\documentclass[aps,pre,superscriptaddress,preprint,12pt,authoryear]{revtex4-1}
\pdfoutput=1
\usepackage{lineno}
\usepackage[latin1]{inputenc}
\usepackage{graphicx}
\usepackage{dcolumn}
\usepackage{bm}
\usepackage{epstopdf}
\usepackage{natbib}
\usepackage{textcomp}

\begin{document}

\title{ Signatures of fluid and kinetic properties in the energy distributions of multicharged Ta ions from nanosecond laser-heated plasma}

 \author{F.Gobet}
\email{gobet@cenbg.in2p3.fr}
\affiliation{Universit\'e Bordeaux, CNRS-IN2P3, CENBG, F-33175 Gradignan, France}
\author{M.Comet}
\affiliation{CEA, DAM-DIF, F-91297 Arpajon, France}
\author{J.-R.Marqu\`es}
\affiliation{LULI-CNRS, \'Ecole Polytechnique, CEA, Universit\'e Paris-Saclay, Sorbonne Universit\'es, F-91128 Palaiseau cedex, France}
\author{V.M\'eot }
\affiliation{CEA, DAM-DIF, F-91297 Arpajon, France}
\author{X.Raymond }
\affiliation{Universit\'e Bordeaux, CNRS-IN2P3, CENBG, F-33175 Gradignan, France}
\author{M.Versteegen}
\affiliation{Universit\'e Bordeaux, CNRS-IN2P3, CENBG, F-33175 Gradignan, France}
\author{J.-L.Henares }
\affiliation{Universit\'e Bordeaux, CNRS-IN2P3, CENBG, F-33175 Gradignan, France}
\author{O.Morice }
\affiliation{CEA, DAM-DIF, F-91297 Arpajon, France}


\date{\today}

\begin{abstract}
The energy distributions of Ta ions produced in a nanosecond laser-heated plasma at 4$\times$10$^{15}$ W.cm$^{-2}$ are  experimentally and theoretically investigated. They are measured far from the target with an electrostatic spectrometer and charge collectors. Shadowgraphy and interferometry are used to characterize the plasma dynamics in the first nanoseconds of the plasma expansion for electron densities ranging from 10$^{18}$ to 10$^{20}$ e.cm$^{-3}$. The experimental data clearly show two components in the energy distributions which depend on the ion charge states. These components are discussed in light of fluid and kinetic descriptions of the expanding plasma.  In particular, quantitative comparisons with calculations performed with 3D hydrodynamic (Troll) and 1D3V Particle In Cell (XooPIC) codes demonstrate that a double layer created at the plasma-vacuum interface plays a crucial role in the acceleration of the highest charge state ions at high energy.

\end{abstract}


\maketitle

\section{Introduction}

High charge state and high energy ions produced in nanosecond laser-heated plasmas are of interest because of their potential applications in thin film deposition \cite{easonpulsed2007}, ion implantation \cite{giuffrida2010surface,wolowski2007application}, ion accelerators \cite{bulanov2002feasibility} or nanoparticle production \cite{donnelly2006pulsed}. The transport of these ions required for such applications depends upon the knowledge of the ion charge state and energy distributions at a distance from the target where recombination processes are stopped and charge states are frozen \cite{roudskoy1996general,burdt2010recombination,burdt2010laser}.

 We are interested in the characterization of ions far from the target in the context of studies of nuclear excitation processes in hot dense plasmas \cite{morel2010calculations,gobet2011nuclear,gobet2008particle,gosselin2007modified}. Theoretical works predict that millions of $^{201}$Hg nuclei could be excited to their first excited state lying 1.565 keV above the ground state in a plasma produced with a 100J/ns laser beam focused at 10$^{15}$ W.cm$^{-2}$ \cite{morel2010calculations}. The detection of these excited nuclei could be possible far from the target as the excited state lifetime is enhanced from 81 ns in neutral atoms to several $\mu$s in ions with charge states beyond 30+ \cite{gosselin2007modified}. The internal conversion, which is the main decay process of this nuclear excited state, is indeed then strongly inhibited. Therefore, the knowledge of the high charge state ion paths from the ion production to their detection is mandatory to estimate the number of excited $^{201}$Hg nuclei that can be detected far from the target. In this study, we have investigated  the energy distributions of Ta ions produced in a laser-heated plasma at 4$\times$10$^{15}$W.cm$^{-2}$, the atomic numbers of mercury and tantalum elements being close.

One possible mechanism responsible for the acceleration of the highest charge state ions in nanosecond laser-heated plasma is the formation of an electric field at the plasma frontier with vacuum, where plasma density is low and in which charge separation between electrons and ions may occur \cite{gurevich1966self,crow1975expansion,eliezer1989double,bulgakova2000double,apinaniz2011theoretical}. This particular region is called the double layer.  The plasma is then divided into two regions: the main bulk, where ions are accelerated under ``thermal"  processes induced by temperature gradients and the double layer, where ions are additionally accelerated under ``electric'' processes. From a theoretical point of view, fluid or kinetic approach are used to describe plasmas. While the fluid description proceeds by numerically solving  the magnetohydrodynamic equations of the plasma, assuming approximate transport coefficients linked to density, pressure or temperature heterogeneities, the kinetic approach considers detailed models of the plasma involving particle interactions in the electromagnetic field. 
One of the key parameters that helps decide which is the best approach for a plasma at an electron temperature $T_e$ and a density $n_e$ is the dimensionless Knudsen number:
$K_n=\lambda/L$
where $\lambda(cm)= 744\frac{T_e^2(K)}{n_e(cm^{-3})}$ is the mean free path of the electrons and $L$ is the plasma gradient defined as $\left|\frac{n_e}{\nabla n_e}\right|$. 
The Debye length $\lambda_D=\sqrt{\frac{\epsilon_0 k_B T_e}{n_e e^2}}$ helps decide whether the plasma is neutral or not.
In high density plasmas for which the Knudsen number is small ($\it i.e.$ $<$ 1) and the Debye length much shorter than the plasma gradient, the plasma can be considered as neutral and the usual continuum approach used in fluid mechanics and heat transfer can be applied. However when the plasma is tenuous, $K_n$ values can become larger than 10 and the Debye length can be of the same order as the plasma gradient. In such a case, the plasma is non-neutral, collisionless and a double layer can be created: a kinetic approach is required. 

Both approaches are considered in the following and quantitatively discussed in relation to a ``thermal'' and an ``electric'' component in the energy distributions of Ta ions measured in their asymptotic state with an electrostatic spectrometer and charge collectors. Interferometry characterizes the plasma dynamics during the 2-3 first ns of the expansion and is used to extract the plasma gradient, which is mandatory to discriminate between the fluid and kinetic approaches. To the best of our knowledge, the quantitative comparison between experimental results on the ion energy distributions with calculations performed with hydrodynamic and kinetic codes is carried out here for the first time. This comparison demonstrates in particular the role of a double layer formation in the acceleration of high charge states at the highest energies in a nanosecond laser heated plasma.

\section{Set-up}

The experimental setup is shown in Fig.1(a). The ELFIE facility at Laboratoire pour l$^{\prime}$Utilisation des Lasers Intenses (LULI) delivers s-polarized pulses with a central wavelength of 1057 nm, of about 35 J and 600 ps duration at half maximum. The main laser beam is focused to a 20 $\mu$m (FWHM) spot onto a tantalum target at an incident angle of 45\r{}, reaching a peak intensity of 4$\times$10$^{15}$ W.cm$^{-2}$. The tantalum target is placed in a vacuum chamber at 10$^{-5}$ mb. It consists of a 2 mm thick slab with a surface area of 1$\times$5 cm$^{2}$. This target is moved by 5 mm after each laser shot to allow shot-to-shot reproducibility. A sub-ps and frequency doubled laser pulse propagating parallel to the target surface with a delay spanning a range of 2 ns before and after the main pulse maximum is used to study the plasma expansion dynamics. The synchronization between the two laser beams was ensured at the focal position with a fast photodiode with an accuracy of $\pm$ 300 ps. This probe beam is combined with a Wollaston crystal, two polarizers and an imaging lens to measure the plasma interference pattern from which the plasma density profile is extracted. Figure 1(b) presents an example of the raw image at the laser main pulse maximum ($t=0$).  Plasma is already well expanded and the threshold between shadowgraphy and interferometry patterns is at 70 $\mu$m from the initial position of the target surface.

\begin{figure}
    \centering
    \begin{minipage}[b]{0.45\linewidth}
        \centering
        \includegraphics[height=7.cm]{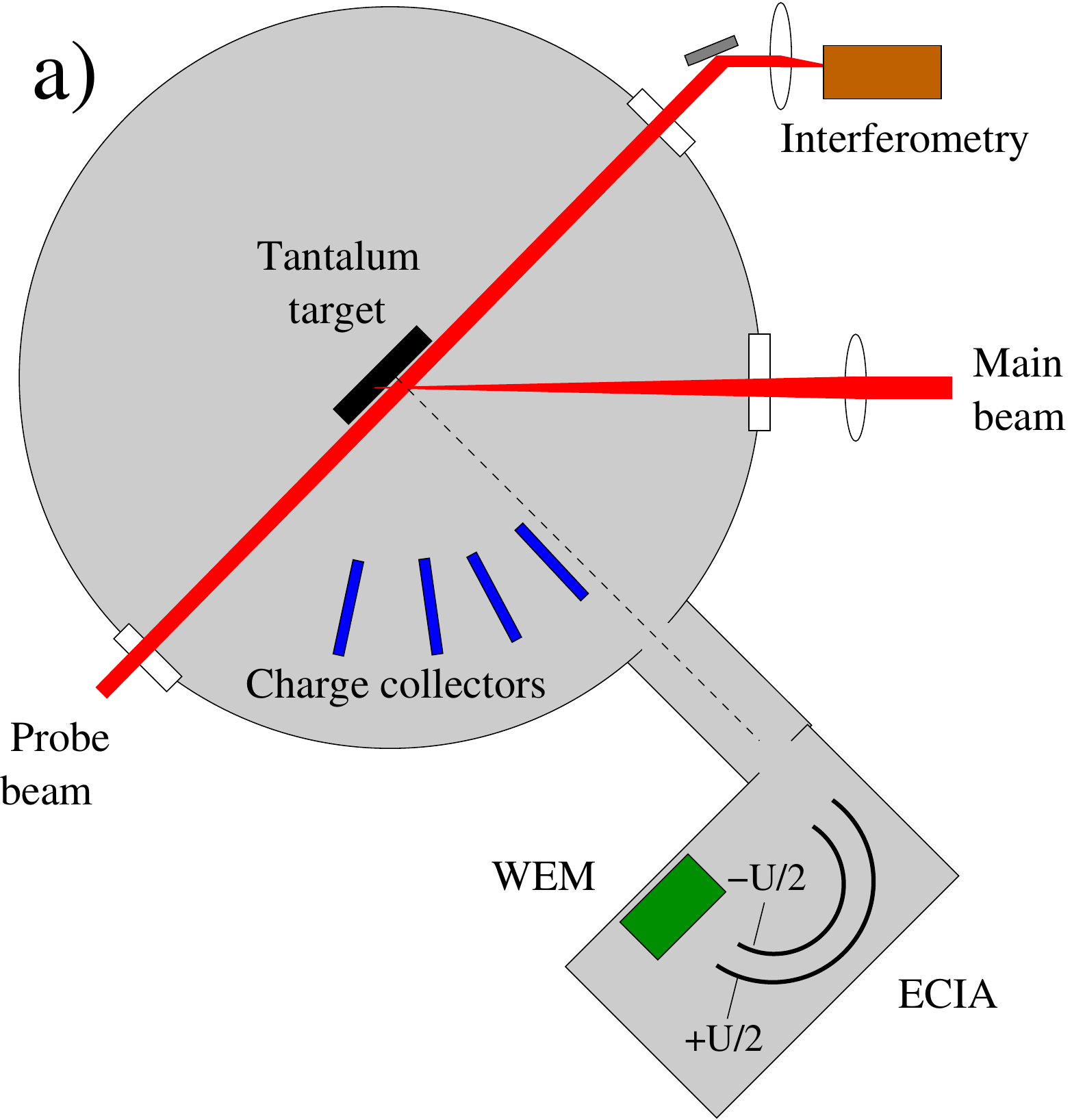}\vspace{0.3cm}\par
        \includegraphics[width=7.cm,trim=0 20 0 90,clip=true]{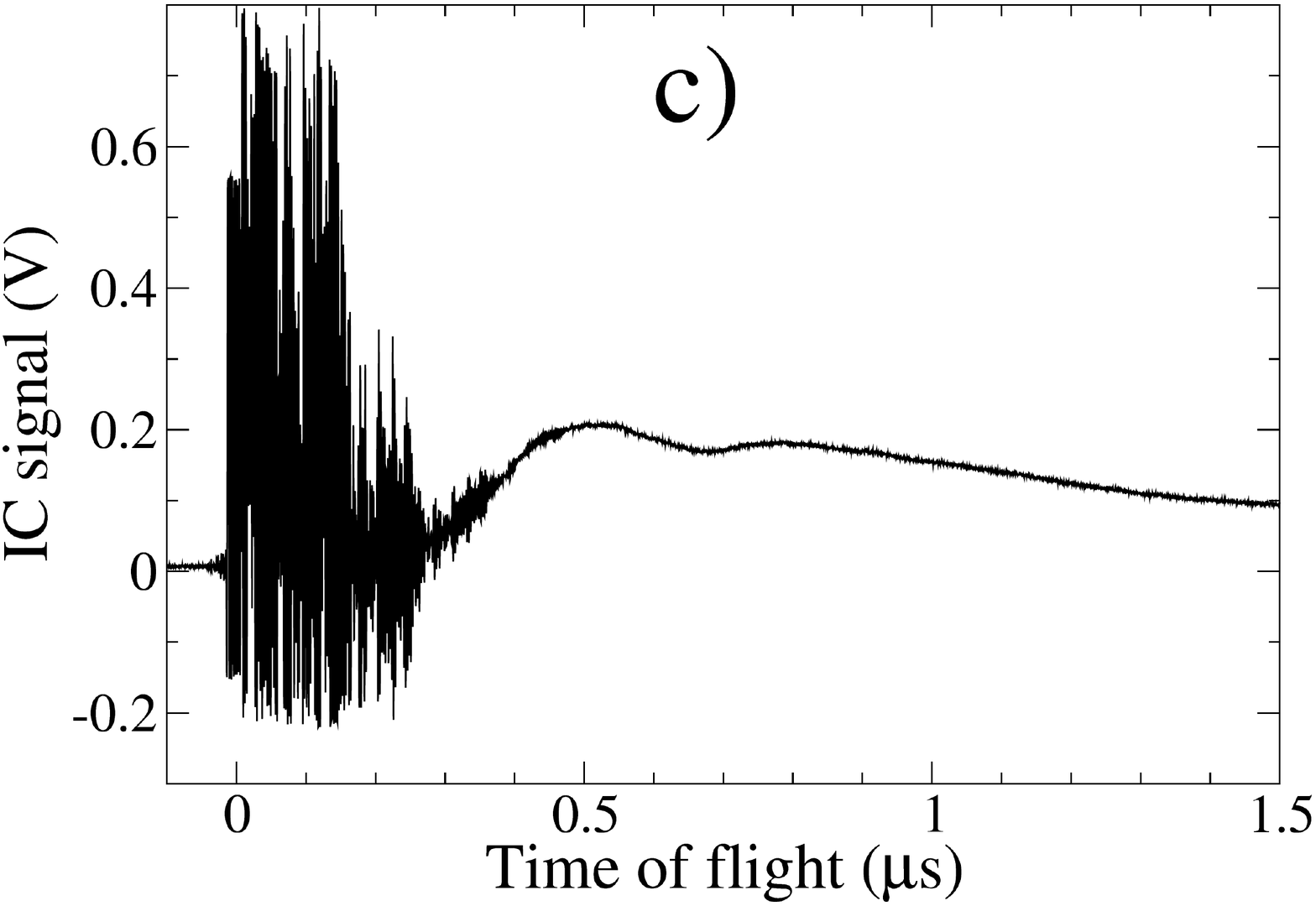}
    \end{minipage}
    \begin{minipage}[b]{0.45\linewidth}
        \centering
	\includegraphics[height=6.cm,trim=0 0 0 0,clip=true]{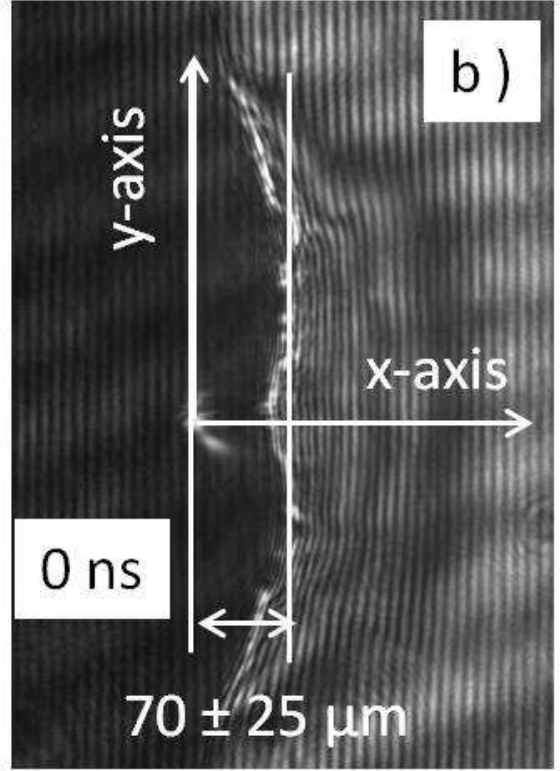}\par
	\vspace{1.3cm}
        \includegraphics[width=7.cm,trim=0 20 0 90,clip=true]{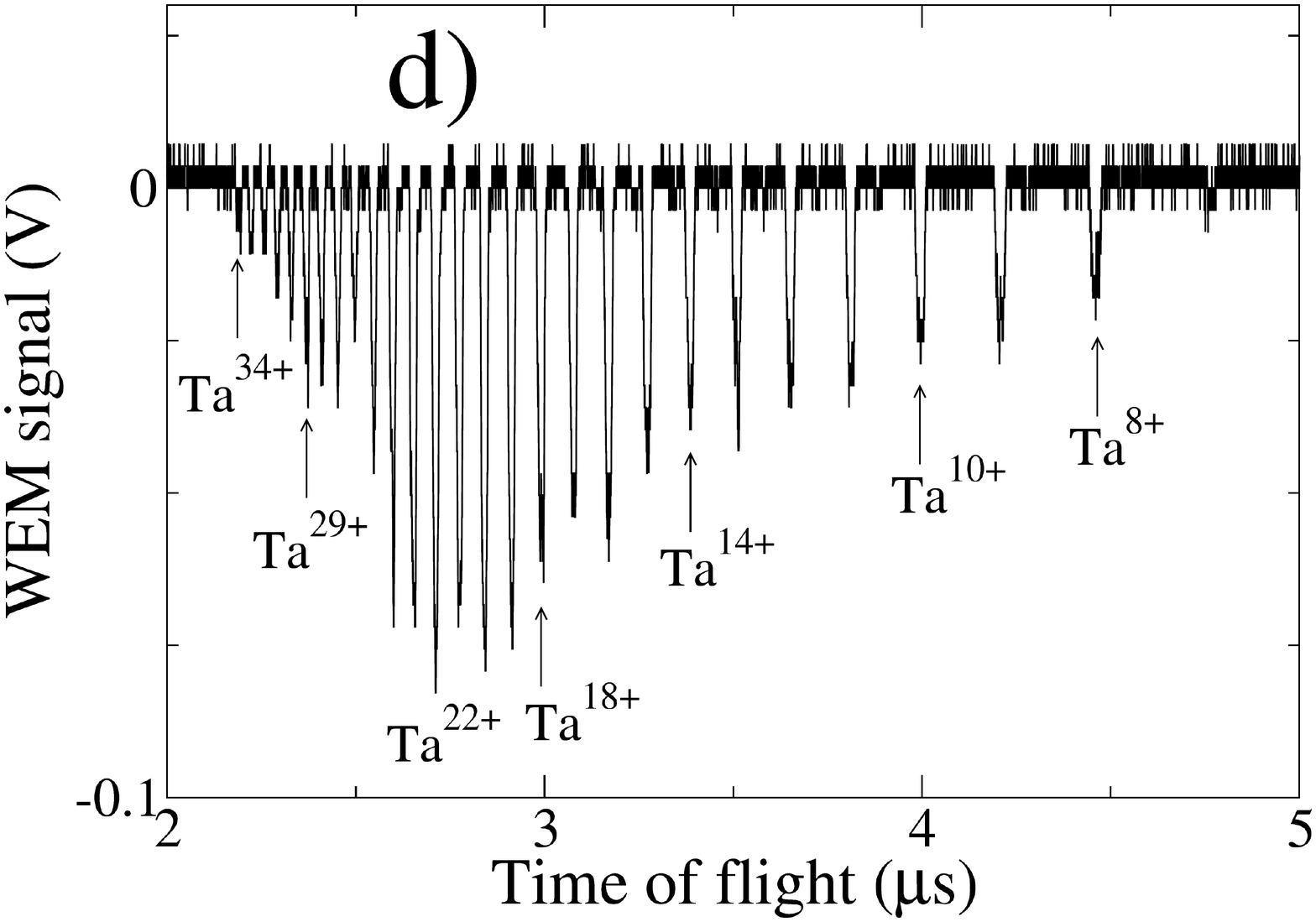}
    \end{minipage}
    \caption{\label{fig:fig1}  (a) Scheme of the experimental set-up. (b) Image of the plasma expansion at the laser main pulse maximum. (c) Ion collector (IC) signal of Ta ions. (d) WEM signal of Ta ions with $U$ = 1100 V. }
\end{figure}

The ions are characterized far from the target with Ion charge Collectors (IC) and an Electrostatic Cylindrical Ion Analyzer (ECIA) coupled with a Windowless Electron Multiplier (WEM). ICs are composed of two biased 3 cm long brass cylinders nested in one another with a drift area allowing charge separation and detection of ions only. The entrance of the outer cylinder is collimated down to a hole 300 $\mu$m in radius. The current generated by the ions collected in the inner cylinder is measured into the 50 $\Omega$ load of a digital oscilloscope which is synchronized with the laser shot. A charge collector  set 30 cm away from the target and at an angle of 14\r{} from the target normal is used as a reference. An example of a time of flight signal from the reference IC is shown in Fig.1(c). The traces of 20 laser shots performed at the same laser intensity are averaged. The signal maximum occuring at $t \sim$ 0.5 $\mu$s is associated with Ta ions with kinetic energies of about 340 keV. A high electric noise induced by laser matter interaction within the first 340 ns hinders the characterization of ions with kinetic energies higher than 750 keV with this diagnostic. Three additional collectors were set 30 cm from the target with incident angles from 0\r{} to 50\r{} to estimate the plasma plume opening angle. A half opening angle of 35\r{} was measured in a range of laser intensities between 2$\times$10$^{14}$ and 10$^{15}$ W.cm$^{-2}$. 


The ECIA used to determine the ion energy distributions has a deflection angle of 180\r{} and is placed in a second chamber 1.8 m from the target in the normal direction. The ECIA chamber is isolated from the target chamber by a differential pumping device and is pumped at 10$^{-6}$ mb. The energy distributions  are measured by biasing the two coaxial metallic cylindrical plates of the ECIA at a voltage $\pm U/2$. Ions with a given energy-to-charge state ratio $E/Q$ exit the ECIA if:
\begin{eqnarray}
 \frac{E(eV)}{Q} = 19.25\   U(V).
\end{eqnarray}
These ions impinge on a calibrated WEM \cite{comet2016absolute} and time-of-flight allows the identification of the ion charge state $Q$. A typical example of a WEM signal is reported on Fig.1(d) at $U= 1100$ V. The signal shows peaks characteristic of charge states of tantalum ions between $Q$=7+ and $Q$=34+.  The areas of these peaks are related to the number of ions selected by the ECIA and impinging on the WEM. These areas, when corrected from the WEM response function and the ECIA transmission, are used as decribed in Ref.\cite{comet2016absolute} to determine the energy distributions of the ions of each charge state.

\section{Experimental results}

Interferometry pictures were recorded for each and every shot. The phase shift, obtained by comparing the plasma fringe position with the background one, is used to extract the density profile by Abel inversion. The electron density profile is consistent from shot-to-shot under the same laser conditions. Figure 2 shows a typical result of the interferometry diagnostic. The top image is a two-dimensional phase profile retrieved from the interferogram (Fig. 1(b)). The plasma density profile is retrieved by Abel inversion. This process
necessarily assumes that the plasma is axisymmetric about the x-axis, so that
only the top half or bottom half of the phase profile are necessary to extract
the density everywhere. However, the code used to analyze the data considers
each half of the phase profile separately, and a smooth transition from one half
to the other increases confidence in the assumption. The bottom graph is a line-out of the longitudinal density profile at  y=100 $\mu$m, close to the x-symmetric axis. The electron gradient length is more than 100 $\mu$m at the pulse maximum. The density threshold between shadowgraphy and interferometry patterns is about 10$^{20}$ e.cm$^{-3}$.
The plasma plume diameter at this electron density is close to 300 $\mu$m and is much larger than the laser focal spot size. This can be explained by strong pressure inside the plume, which is sufficient to push matter radially. In such a situation, ions are emitted in a wide cone angle \cite{bulgakova2000double}. This result is qualitatively in agreement with the half opening angle of 35\r{} measured with the array of ion collectors.

\begin{figure}
    \centering
    \begin{minipage}[b]{0.7\linewidth}
     \centering
    \hspace{0.5cm}
     \includegraphics[width=6.cm,trim=0 0 0 0,clip=true]{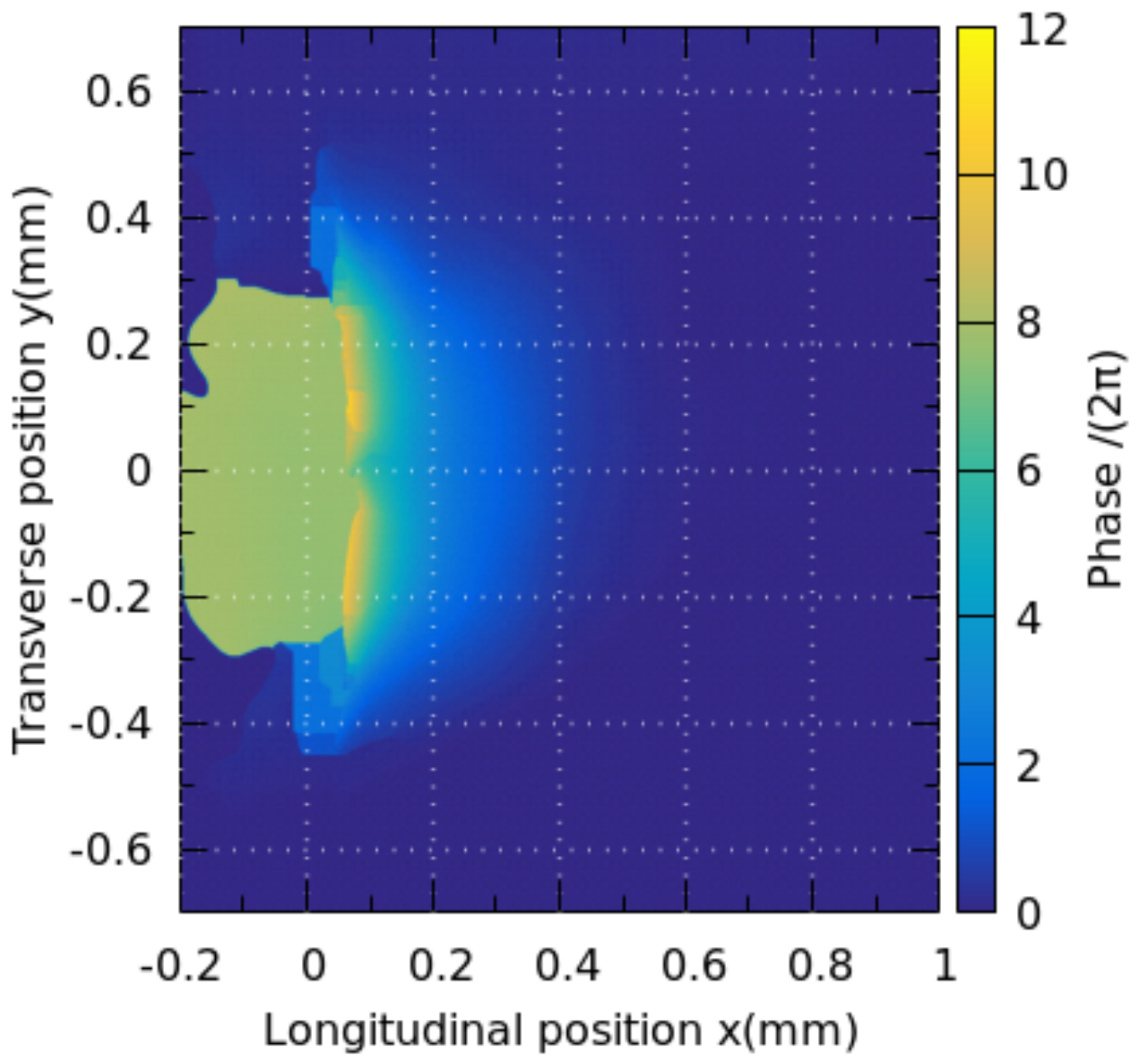}\par
     \includegraphics[width=6.cm,trim=16 40 340 210,clip=true]{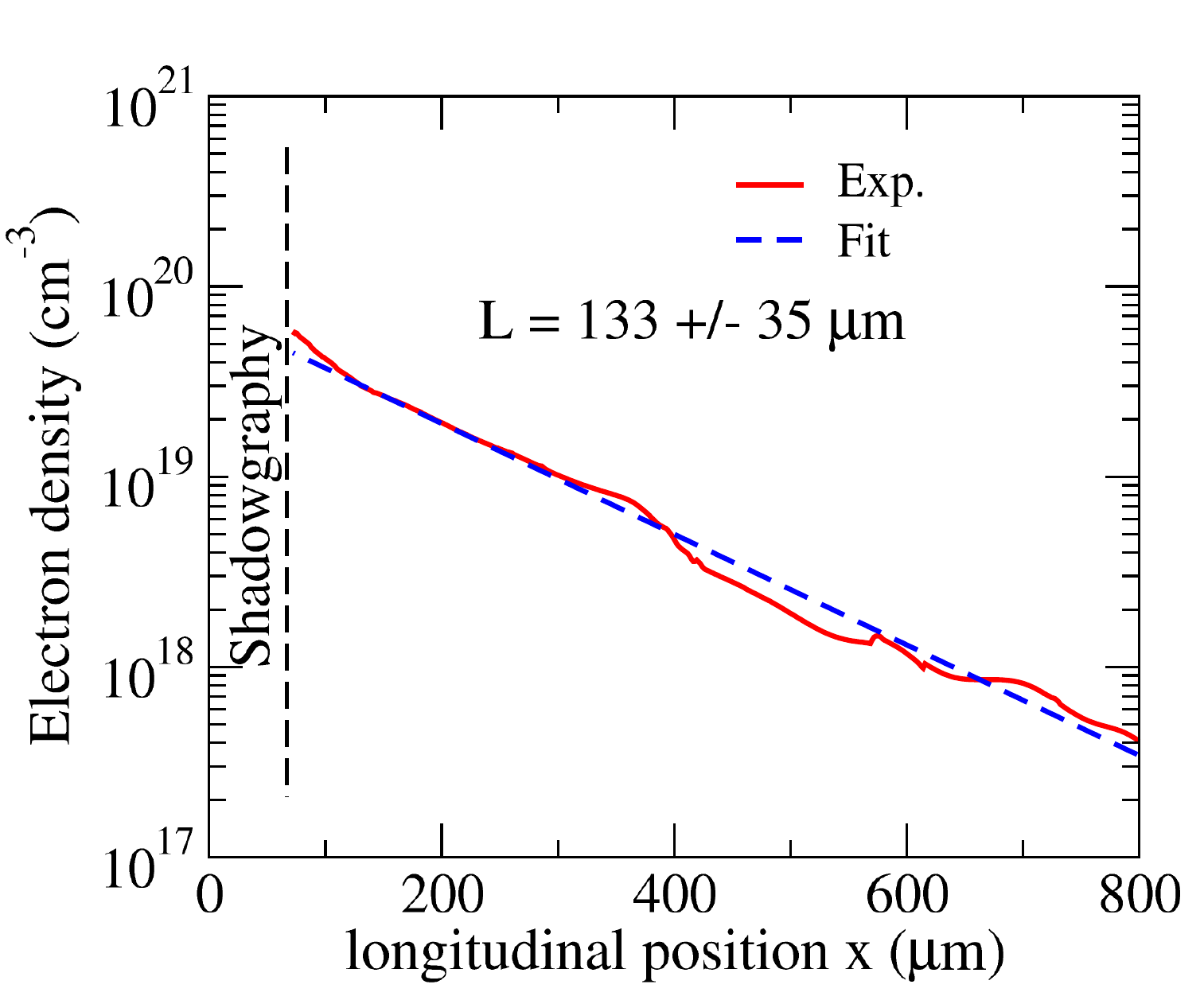}
    \end{minipage}
    \caption{\label{fig:fig2}  Plasma expansion at the instant of the laser main pulse maximum ($t=0$). Top: Phase shift pattern. Bottom: Electron density profile. The longitudinal position x=0 is the position of the target surface before the shot.}
\end{figure}


The energy distribution of the total ion charge can be obtained from the IC signal $i(t)$ using the following transformation:
\begin{eqnarray}
 \frac{d^2Q}{dEd\Omega}(E=m_{Ta}l^2/2t^2) = \frac{i(t) t^3}{m_{Ta}S}
\end{eqnarray}
with $m_{Ta}$ the tantalum ion mass, $l$ the distance between the target and the IC entrance,  $t$ the time of flight and $S$ the IC entrance area. The measured charge averaged over bins of 20 keV is reported in Fig. 3. As shown in Fig. 1(c), the time of flight traces have poor signal-to-noise ratio below 340 ns and the charge energy distribution is truncated above 750 keV. The charge exhibits two decreasing exponential distributions with strong slope modification around 150 keV. This behavior already signs two different ion populations.
\begin{figure}[t]
{\includegraphics[width=7cm,trim=0 0 130 80,clip=true]{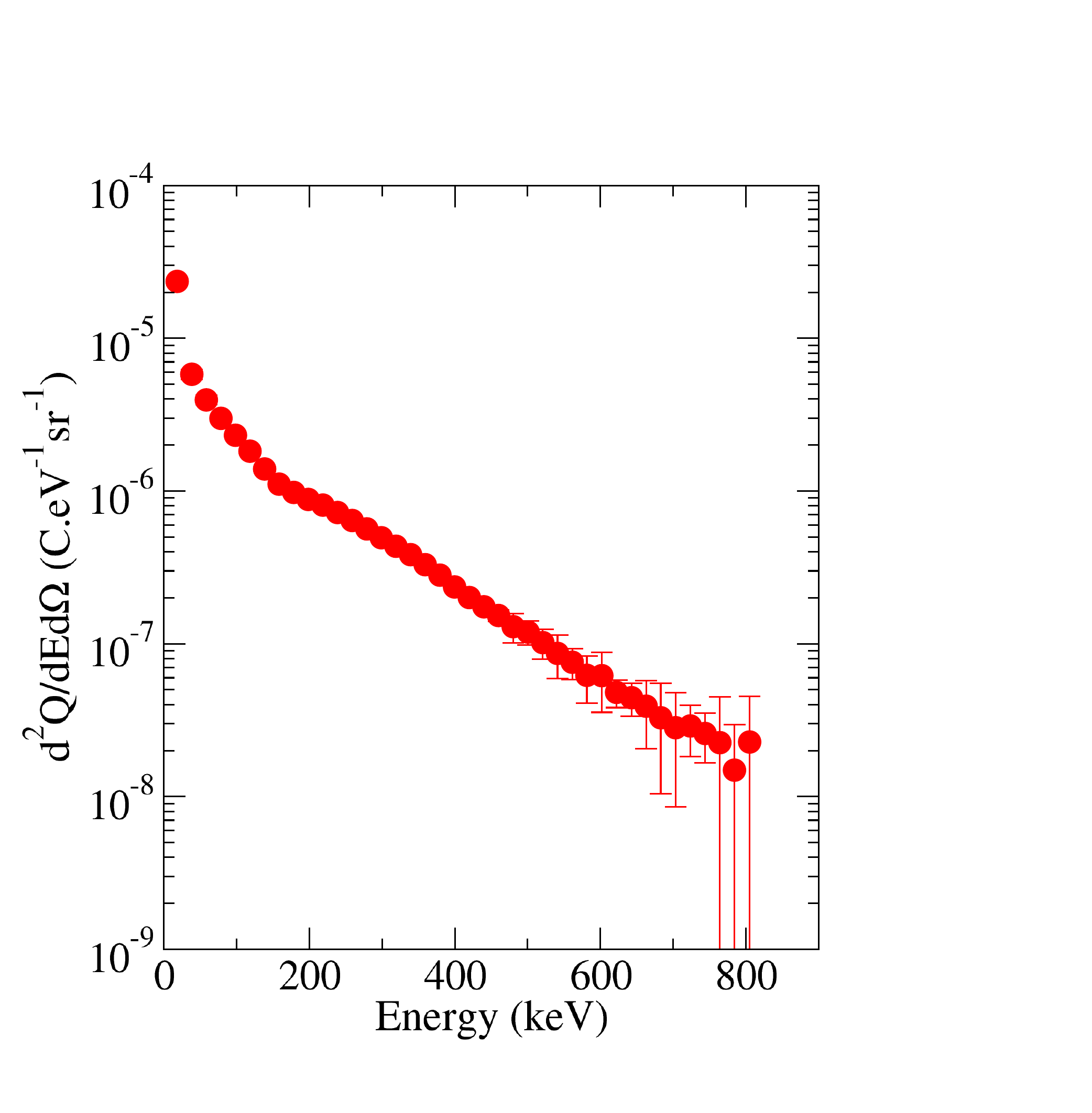}}
\caption{\label{fig:fig3}  Energy distribution of the Ta ion charge}
\end{figure}

The absolute energy distributions of each tantalum plasma ion have been measured using the ECIA for charge states between 1+ and 37+. The WEM response, measured in Ref.\cite{comet2016absolute} for tantalum ion energies lower than 30 keV, has been extended to higher energies taking the energy dependence determined by Krasa et al. into account \cite{krasa1998gain}. The ECIA transmission has been estimated following the protocol described in Ref.\cite{comet2016absolute}. It consists in comparing the measured IC signal with the reconstructed IC signal built from the ECIA data under the hypothesis that the ion trajectories in the ECIA are unaffected by space charge effects. With this method the ECIA transmission has been estimated to be of the order of 100 \% for ion energies higher than 350 keV and reaches a minimum of 20 \% at 70 keV. Figure 4(a) shows the energy distributions for the 4+, 9+, 13+, 20+, 25+ and 30+ charge states. The distributions are continuous with a regular shift of the maximum towards higher energies with higher charge states.

The corresponding velocity distributions have been shown to follow the phenomenological shifted Maxwell-Boltzmann-Coulomb (MBC) distributions \cite{kelly1992gas,sibold1991kinetic,miotello1999origin,kelly1990dual,krasa2013gaussian,torrisi2006energy} that empirically take into account two main properties to describe the laser-heated plasma expansion in vacuum: a thermal component and a high electric field related to charge density heterogeneity which accelerates ions to higher velocities. As a consequence the target normal component of the velocity distributions is generally shifted by a velocity value containing two terms: a plasma center of mass velocity corresponding to its adiabatic expansion stage in the Knudsen layer $v_T$ \cite{kelly1992gas,sibold1991kinetic} and a Coulomb interaction velocity $v_C$ \cite{torrisi2006energy}. The expressions of these two phenomenological components are given by:
\begin{eqnarray}
v_T=\sqrt{\frac{\gamma kT}{m_{ion}}} ,\quad v_C=\sqrt{\frac{2eQV_0}{m_{ion}}}
\end{eqnarray}
where $k$ is the Boltzmann constant, $T$ is the effective ion temperature, which is a way to evaluate the spread of the velocity distribution around the center-of-mass velocity, $\gamma$ is the adiabatic coefficient ($\gamma$=$\frac{5}{3}$ in our case of monoatomic species), and $V_0$ is the equivalent accelerating voltage developed in the plasma and which accelerates the ions to an energy proportional to their charge state in the direction normal to the target. The corresponding energy distributions of ions entering the detector with kinetic energies between $E$ and $E$+$dE$ are given by \cite{comet2016absolute}:
\begin{eqnarray}
\frac{d^2N^Q}{dEd\Omega}(E)=C_Q\ E\ exp\left[-\frac{E+E_{MBC}-2\sqrt{EE_{MBC}}}{kT}\right]
\end{eqnarray}
where $N^Q$ is the number of ions of charge $Q$, $C_Q$ is a proportionality coefficient, and $E_{MBC}=\frac{1}{2}m_{ion}(v_T+v_C)^2$. The MBC function of Eq.(4) is used to fit the data.
The fitting parameters are the number of ions $C_Q$, the effective ion temperature $T$ and the equivalent voltage $V_0$. Figure 4(b) shows examples of $\chi^2$ values obtained for different ($T$,$V_0$) couples and different tantalum ion charge states $Q$. The lowest $\chi^2$ values are of the order of 0.1 showing an overestimation of the statistical error bars on experimental data. Nevertheless, the $\chi^2$ values clearly show that for low charge states the MBC functions that fit the distributions have a small and compatible with zero effective potential, whereas for high charge states an effective potential of about 15 kV is necessary. The good agreement between experimental data and fits reported in Fig.4(a) with this phenomenological model demonstrates that the ions exhibit two different populations depending on their charge states, which we will talk the ``thermal component'' for ions with charge states lower than 13+ and the ``electric component'' for ions with higher charge states and energies larger than 300-400 keV.

\begin{figure}
    \centering
    \begin{minipage}[b]{0.4\linewidth}
     \centering
     \includegraphics[height=8.cm,trim=0 60 352 0,clip=true]{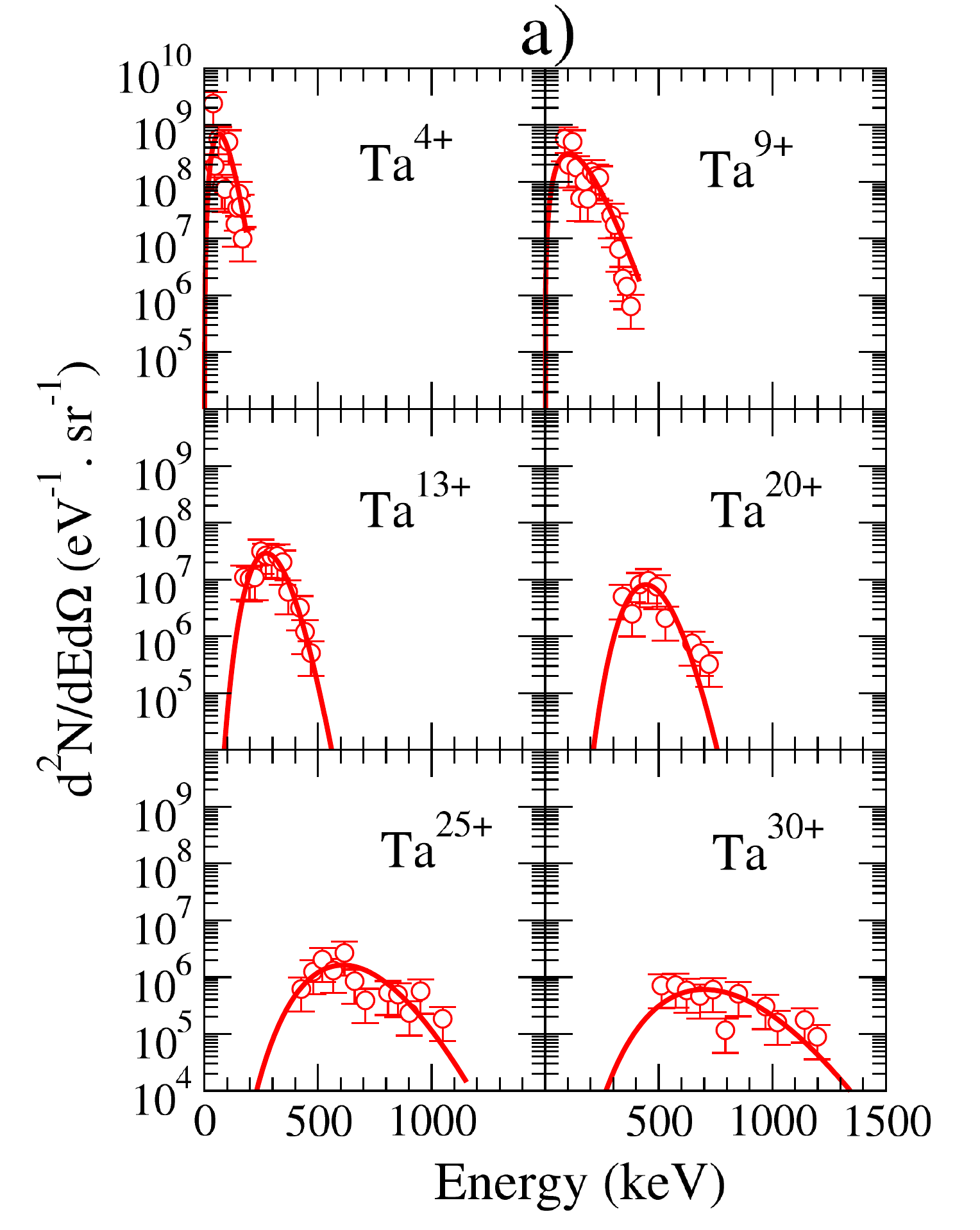}
    \end{minipage}
    \begin{minipage}[b]{0.2\linewidth}
    \centering
        \includegraphics[height=8.cm,trim=0 0 0 10,clip=true]{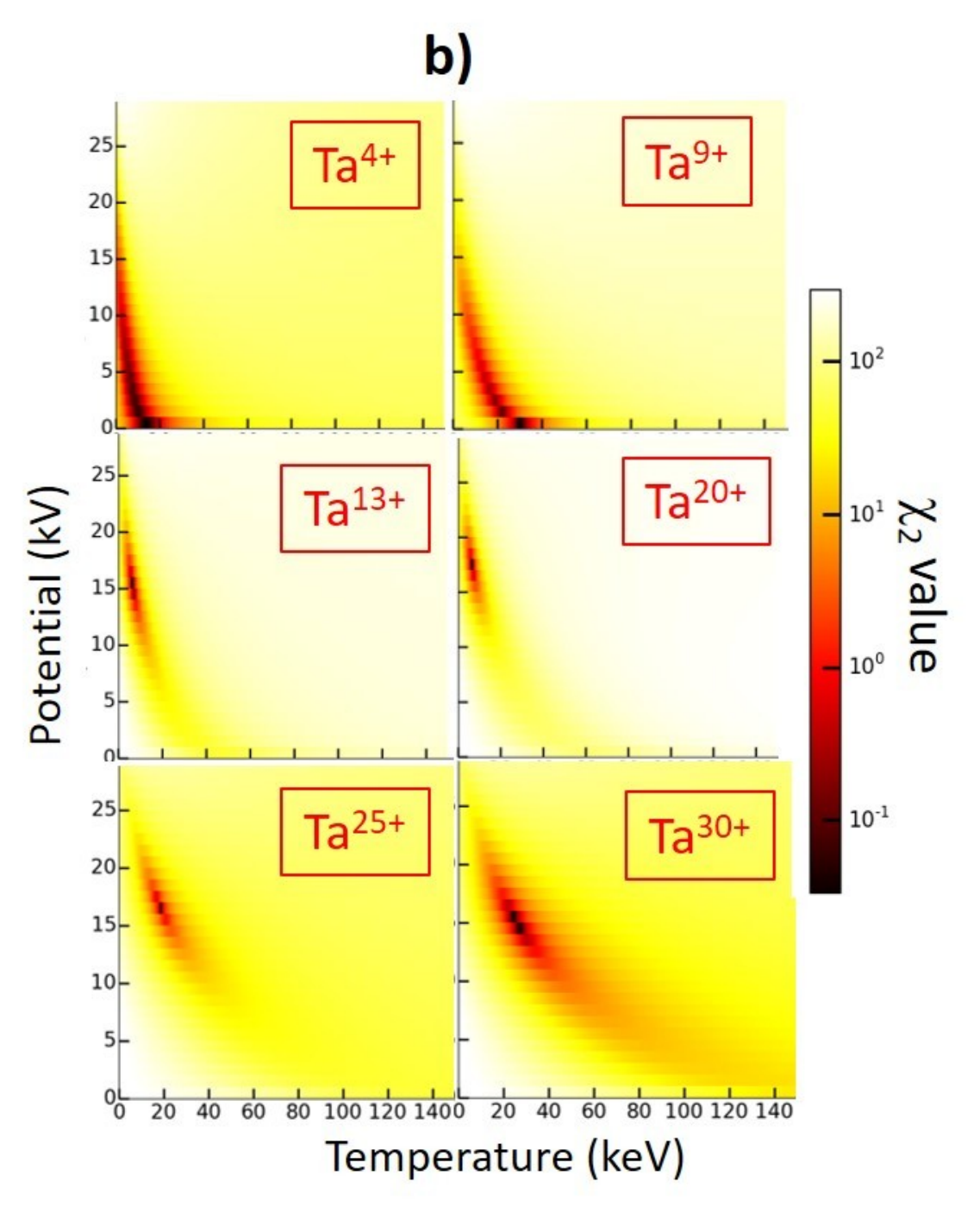}\vspace{1.mm}
    \end{minipage}
\hspace{3cm}
    \caption{\label{fig:fig4}  (a) Measured kinetic energy distributions of Ta ions produced at a laser intensity of 4$\times$10$^{15}$ W.cm$^{-2}$ and the corresponding MBC function fits. (b) $\chi^2$ values between experimental data and MBC functions for different ($T$,$V_0$) couples.}
\end{figure}

Figure 5 shows the evolution of the ion yield as a function of their charge state. This yield is obtained by integrating the MBC functions from 0 to 1500 keV. Considering a solid angle of about 1 sr in regards to the 35\r{} half opening angle, the number of ions in the ``electric'' component with $Q\ge$15+ is about 3-5 10$^{12}$ ions and only represents about 5\textperthousand$ $ of the total number of ions emitted in the two components. Therefore the electric field responsible for additional acceleration of the highest charge state ions does not occur in the whole plasma volume. Ions in the ``electric'' component are produced in very specific plasma regions and follow particular paths from their production to their detection. 

\begin{figure}[t]
{\includegraphics[width=6.4cm,trim=0 41 340 70,clip=true]{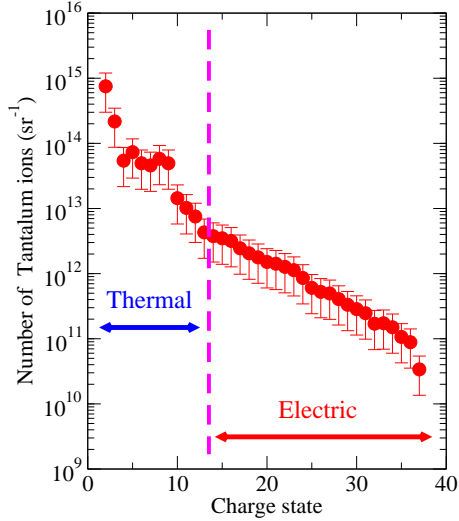}}
\caption{\label{fig:fig5} Ta ion yields as a function of their charge state.}
\end{figure}

The model upon which the MBC distributions are based is phenomenological. Other models have been developped to study the acceleration of high charge state ions in plasmas heated by nanosecond laser pulses. They have been discussed in different works in the framework of charge separation in the periphery of the expanding plasma \cite{gurevich1966self,crow1975expansion,eliezer1989double,bulgakova2000double,apinaniz2011theoretical}. In this particular region, the electron cloud overtakes the ions leading to the formation of a double layer and an electric field. This scenario is similar to the one of ion acceleration in the Target Normal Sheath Acceleration regime with sub-ps laser pulses on solid targets. This regime has been extensively studied in the framework of ultra-short laser pulses with many theoretical works under different assumptions, such as isothermal or adiabatic expansion or non-Maxwellian distributions of electrons (see the review article of Macchi et al. \cite{macchi2013ion} and references therein). In Mora$^\prime$s work \cite{mora2003plasma}, the author assumes a plasma initially at rest in the reference frame composed of ions with a single charge state $Q$. Considering Boltzmann distributed electron energies and isothermal expansion of the plasma, Mora provides the following expression for the energy distribution of the ions:
\begin{eqnarray}
\frac{d^2N^Q}{dEd\Omega}(E)\propto \frac{1}{\sqrt{EE_0}}\ exp(-\sqrt{2E/E_{0}})
\end{eqnarray}
where $E_0$ is a characteristic energy dependent on the charge state $Q$ of the ions and the electron temperature $T_e$ as $E_0$=$QkT_e$.

Although the Ta laser-produced plasma in this work contains many possible charge states of the tantalum element, it is interesting to compare the experimental energy distributions of the ions in the ``electric component'' with Mora$^\prime$s function rewritten in the framework of a plasma boundary moving at the velocity $v_B$  under thermal expansion in the laboratory reference frame. The ion energy distribution is found to be:
\begin{eqnarray}
\frac{d^2N^Q}{dEd\Omega}(E)\propto \frac{1}{\sqrt{EE_0}}\ exp\left(-\sqrt{\frac{2E+2E_{B}-4\sqrt{EE_{B}}}{E_{0}}}\right)
\end{eqnarray}
where $E_B$ is the ``kinetic energy'' of the boundary given by $\frac{1}{2}m_{Ta}v_B^2$. This energy is fixed at 400 keV in the following, the detected ions in the ``thermal component'' having energies lower than this value. Figure 6 shows the measured energy distributions of ions with $Q\ge$13+ for which an ``electric component'' appeared in the MBC fit. They are compared with the theoretical predictions discussed above. Mora$^\prime$s model given by Eq.(6) is used to fit the data with $E_0$ as a free parameter. The agreement between fit and data strengthens the scenario of ions accelerated by an electric field produced in a double layer at the frontier between plasma and vacuum. The characteristic ion energy $E_0$ depends on the ion charge state. It is reported in Fig.7 and increases from $\sim$1 keV for $Q$=13+ to about 60 keV for $Q\ge$25+. This behavior is different from the Mora$^\prime$s linear prediction given for plasmas with single charge state ions and fixed electron temperature. 

\begin{figure}[t]
{\includegraphics[width=6.4cm,trim=0 61 360 50,clip=true]{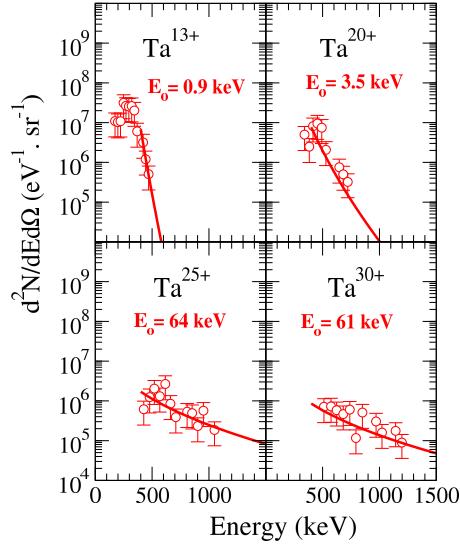}}
\caption{\label{fig:fig6} Measured kinetic energy distributions of Ta ions with the modified Mora$^\prime$s function fits (see text for details).}
\end{figure}

\begin{figure}[t]
{\includegraphics[width=6.4cm,trim=0 44 360 170,clip=true]{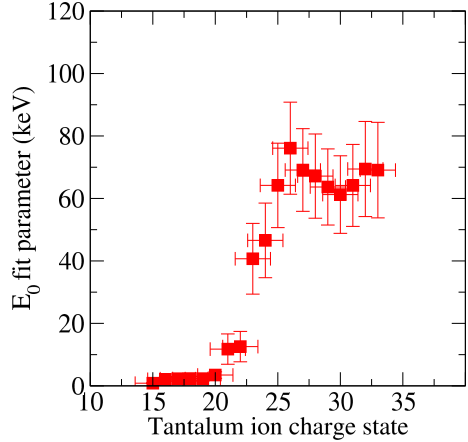}}
\caption{\label{fig:fig6plus} Evolution of $E_0$ fit parameter as a function of the tantalum ion charge state.}
\end{figure}

\section{Fluid and kinetic simulations. Discussions.}

The experimental ion energy distributions and their comparisons with a phenomenological and an analytical model seem to sign at least two different acceleration mechanisms depending on the ion charge state. These observations are here supported using fluid and kinetic descriptions.

\subsection{Fluid description: ``thermal'' component}

Simulation of the plasma dynamics in its densest part has been performed over 3 ns spanned around the laser shot with the radiative-hydrodynamic code Troll \cite{lefebvre2018}. The simulation is performed using a 3D Lagrangian method. The photon transport is modeled using a multigroup Monte-Carlo coupled to fast non-LTE opacity tables \cite{busquet1993radiation,bowen2003dielectronic, bowen2004gold}. The electron heat flux uses a Spitzer-Harm flux limiter of 0.15. The equation of state used in the simulation is relying on a quantum average-atom model calculation as described in \cite{wetta2018}. The laser propagation, refraction and collisional absorption are treated by a ray tracing algorithm. The simulation has been carried out with a laser intensity characterized by the Gaussian temporal shape $I(t)=I_oe^{-4\ ln2\ t^2/\tau^2}$ with pulse duration $\tau$=0.6 ns. The intensity profile on target considered in the code is the one measured experimentally. It has a double Gaussian distribution $I(x,t)=I(t)[0.91 e^{-x^2/R_1^2}+0.09e^{-x^2/R_2^2}]$ with spot radiuses $R_1$ = 10.3 $\mu$m and $R_2$ = 39.9 $\mu$m. The laser beam energy has been set at 35 J and the incident angle fixed at 45\r{}.


Figure 8 presents the electron temperature and density distributions in the incident plane calculated with the Troll code at $t=-0.25$ ns, $ t=0$ ns and $t=0.5$ ns. The laser pulse comes at 45\r{} from the bottom of the figure. At the pulse intensity maximum, the plasma reaches a radial diameter of about 300 $\mu$m in agreement with the experimented value estimated from Fig.1(b). Moreover, the plasma plume is elongated with a radius about two times smaller than its length, similarly to the half open angle of 35\r{} mesured with the IC array. The highest electron temperature reaches 3 keV near critical density, whereas the plasma at the frontier with vacuum experiences a fast cooling from 1600 eV down to 400 eV in 400 ps. The temperature and density distributions exhibit an almost cylindrical symmetry around the target normal despite a 3D geometry of the laser interaction. The focal spot size being relatively small, electronic conduction rapidly smoothes this asymmetry.

To check whether the Abel inversion is efficient to extract the electron density profile, the interferometry diagnostic is included in the simulation. A simulated probe beam passes through the plasma at 90\r{} to the target normal. For each probe beam ray, the phase is modified proportionally to the integral of the electron density along the optical path of the ray. The phase shift $\Phi$ is then calculated by:
\begin{eqnarray}
\Phi = \frac{2\pi}{\lambda} \int\left(\sqrt{1-\frac{n_e(l)}{n_c}} -1\right) dl
\end{eqnarray}
with $\lambda$=528 nm the wavelength of the probe beam, $n_e$ the electron density and $n_c$=3.9$\times$10$^{21}$ e.cm$^{-3}$ the critical density at this wavelength. As in the experimental setup, an imaging system is placed in the simulation to obtain the plasma image. With the simulated phase shift cartography, an Abel inversion is performed to infer the electronic density. A comparison of the electronic density determined by Abel inversion with the one directly extracted at the target normal from the simulation shows a good agreement. This result confirms the validity of the Abel inversion algorithm to retrieve the electron density profiles.

\begin{figure}[t]
{\includegraphics[width=7.33cm]{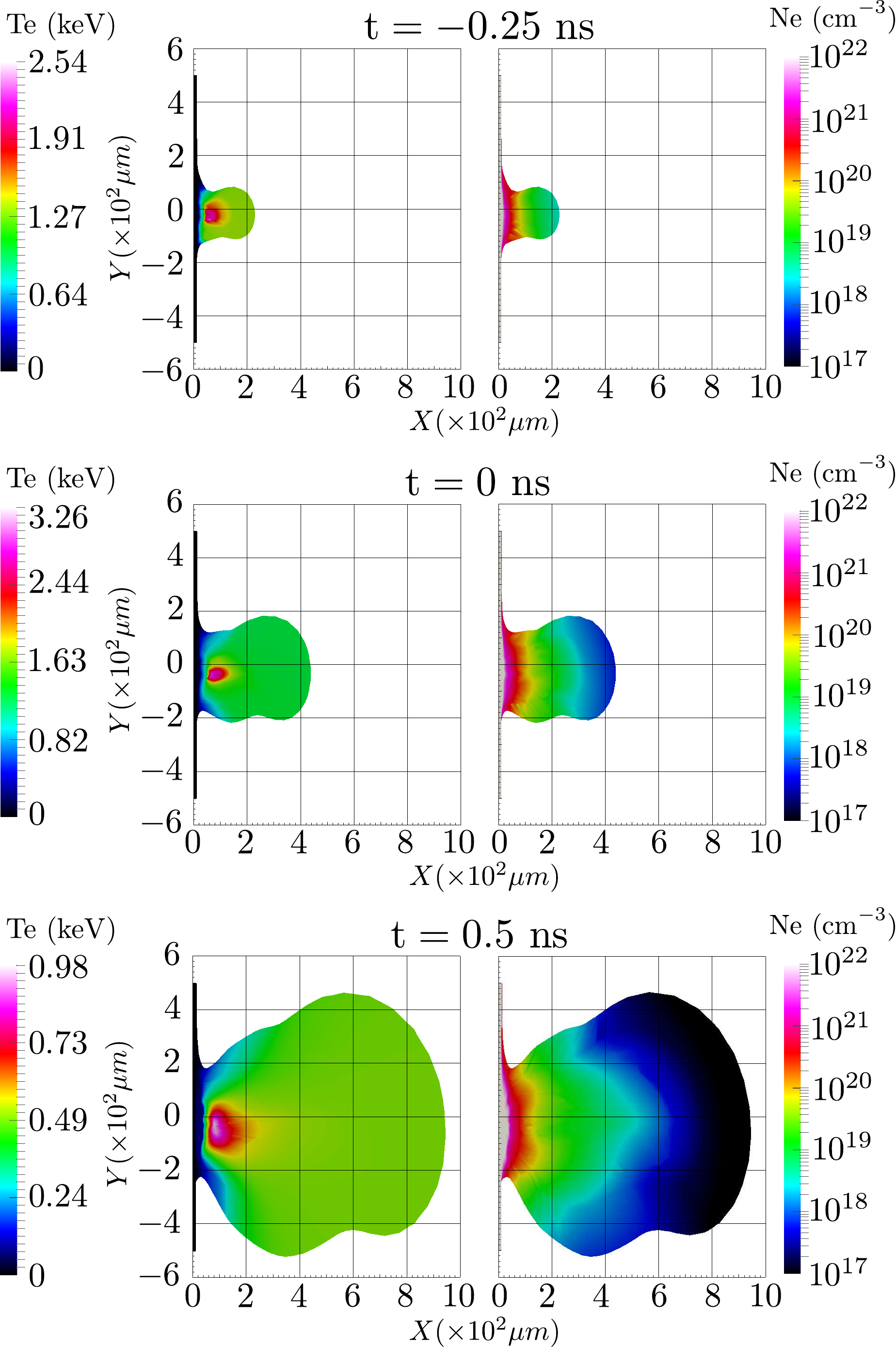}}
\caption{\label{fig:fig6b} Electron temperature and density in the incident plane calculated with the Troll code for a tantalum plasma heated by a nanosecond laser pulse at 4$\times$10$^{15}$ W.cm$^{-2}$ at three different instants, $t=-0.25$ ns, $t=0$ ns, $t=0.5$ ns. The maximum laser intensity is reached at $t=0$ ns. The target is at the longitudinal position x=0.}
\end{figure}

At different instants of the calculation, the electron density profile has been extracted along the target normal from the grid elements verifying the criteria $K_n\le$1. This allows the calculation of the plasma gradient in the density range $10^{18}-10^{20}$  e.cm$^{-3}$ as well as the distance from the initial target surface of the isodensity line at  $10^{20}$  e.cm$^{-3}$. Figure 9 shows the evolution of these observables as a function of time and their comparison with the experimental values. The measured plasma gradient exhibits a maximum about 0.5 ns after the pulse maximum whereas the isodensity line distance increases monotonously up to 2 ns
with a slight rebound at 0.5 ns. A good agreement is observed between experimental and calculated values with a time offset of about 300 - 500 ps. This offset could be due to the systematic error on the synchronization of the main and probe beams with the fast photodiode.  In the first ns, the plasma experiences a fast thermal expansion because of laser heating whereas the rebound observed after 0.5 ns in the shadowgraphy pattern is induced by the mechanical response of matter initially compressed by the ponderomotive force of the laser pulse. The corona follows an adiabatic expansion after the end of the laser pulse leading to the decrease of the gradient observed after 1 ns. 

\begin{figure}[t]
{\includegraphics[width=7.33cm,trim=0 35 230 0,clip=true]{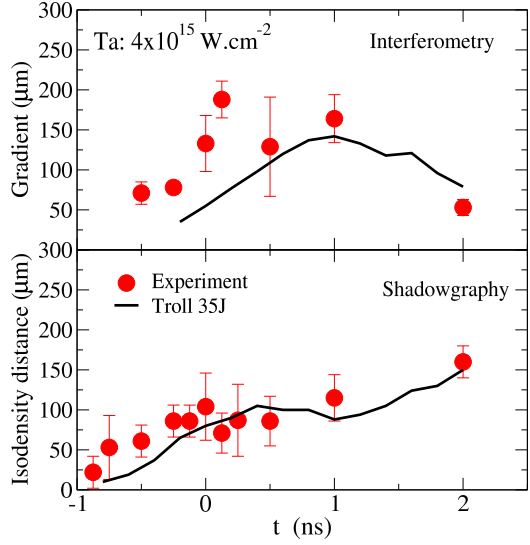}}
\caption{\label{fig:fig7} Comparaison between experimental and calculated gradient (up) and isodensity line distance (bottom) reported as a function of time. The maximum intensity of the main pulse is reached at $t=0$.}
\end{figure}

Calculations show that at the pulse intensity maximum the highest ion temperature reaches around 600 eV near the critical density whereas at the end of the laser pulse, the ion temperature decreases down to about 100 eV. Whatever the instant of the laser and the plasma density, these ion temperatures present low values and the measured ion energies are not representative of the thermal energy. Additional acceleration mechanisms are required to interpret the ion energies as those induced by strong thermal and pressure gradients in the plasma bulks.
Fig.10(a) displays the tantalum ion energy distributions calculated with Troll for a laser energy of 35 J on target at four instants before and after the laser pulse maximum ($t=0$). The distributions are constructed considering ions in all the grid elements verifying the criteria $K_n\le$1. Two behaviors are observed depending on time. The plasma is heated throughout the application of the laser pulse, producing ions with increasing energies up to 500 keV in the fluid approximation. Once the laser pulse ends, plasma adiabatically expends in vacuum and ion energy distributions display a steady shape up to 2 ns. These distributions are compared in Fig.10(b) with the asymptotic ion energy distributions built from the summation at a given energy of all the MBC functions fitting the measured data. A differentiation is made depending on the ion component and therefore on the ion charge state. The green curve represents the total number of ions reported with error bars as the dashed green curves corresponding to a statistical uncertainty of $\pm$ 60\%.  It is worth noticing that the in situ calculated distributions span a range of energies in agreement with the experimental ``thermal'' component of the distributions. Despite strong physical difference between both quantities, $\it i.e.$ ion distributions calculated during the first stages of plasma expansion versus ion distributions measured in frozen states more than 1 $\mu$s after the laser shot, this qualitative agreement suggests that fluid effects are responsible for the acceleration of ions measured at energies lower than 400 keV. These ions constitute the main part of the plasma particles observed far from the target. Moreover, it is clear that energetic ions observed above 1 MeV cannot be interpreted by fluid description and an additional acceleration process occurring somewhere else than the dense part of the plasma is required to interpret the experimental data.

\begin{figure}[h]
{\includegraphics[width=10cm,trim=0 160 180 70,clip=true]{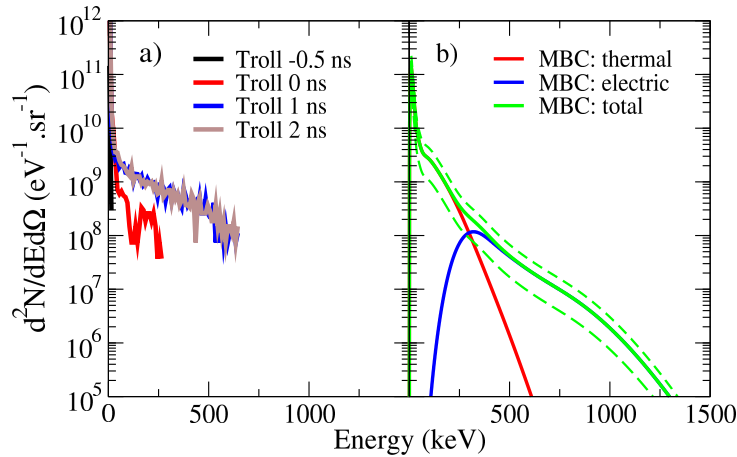}}
\caption{\label{fig:fig8} a)  energy distributions of ions calculated at different instants about the laser pulse maximum ($t=0$ ns) with the Troll code.  b) asymptotic ion energy distributions built from the MBC functions fitting the measured data.}
\end{figure}

\subsection{Kinetic description: ``electric'' component}

A double layer formed at the frontier between plasma and vacuum produces an electric field that could accelerate ions at energies higher than the thermal component values. We will now try to determine whether a double layer can appear under our experimental conditions, and if so, if its characteristics, in terms of electron density and temperature, can be used to reproduce the measured energy distributions of the highest Ta ion charge states.

In order to be efficient for additional acceleration of the highest charge state ions, the typical time for the double layer formation must be smaller than the typical time of the plasma evolution. We will thus first focus on the evolution of the plasma temperature, mean charge state and electron density at the plasma frontier with vacuum as a function of time using the fluid code Troll. Then, we will determine the electron density range over which a double layer can be formed, using 1D3V PIC simulations with inputs from the Troll fluid calculations. Two conditions must be verified: 1) the time needed for the double layer formation must be low enough so that the plasma parameters do not significantly evolve and the acceleration of the highest charge state ions is possible before the plasma cooling stage, 2) the Debye length must be larger than the plasma gradient. These conditions will determine the electron density range over which the double layer is formed. It will finally be used as an input of PIC simulation to extract the energy distributions of ions accelerated in the double layer of the multispecies plasma for comparison with the ``electric'' component observed in the experimental data.

\subsubsection{Double layer formation conditions}

Figure 11 shows the evolution of the ion mean charge state, the electron temperature and density at the frontier between plasma and vacuum, from Troll calculations. The  temperature reaches 1400-1600 eV in the first half of the laser pulse and decreases down to 100 eV in about 1 ns. 
The evolution of the mean charge with time is correlated with temperature and increases from 4+ to 43+ in the first 200 ps of the arrival of the laser pulse, then stays stable at 43+ up to the last 200 ps of the laser pulse when it exhibits a continuous decrease from 43+ down to 14+. Calculations show that the grid element describing this plasma region has a mean electron density of the order of $10^{16}-10^{18}$ e.cm$^{-3}$ at the pulse intensity maximum and during the next two nanoseconds. This grid element also has a longitudinal size of a few hundreds of micrometers.
Strong density heterogeneities should thus occur in this plasma region which shows that the fluid description is not appropriate here. However the order of magnitude of the highest temperature and plasma evolution time will be used in the following to estimate at which electron density a double layer occurs in this plasma region.

\begin{figure}[h]
{\includegraphics[width=7.5cm,trim=30 35 220 0,clip=true]{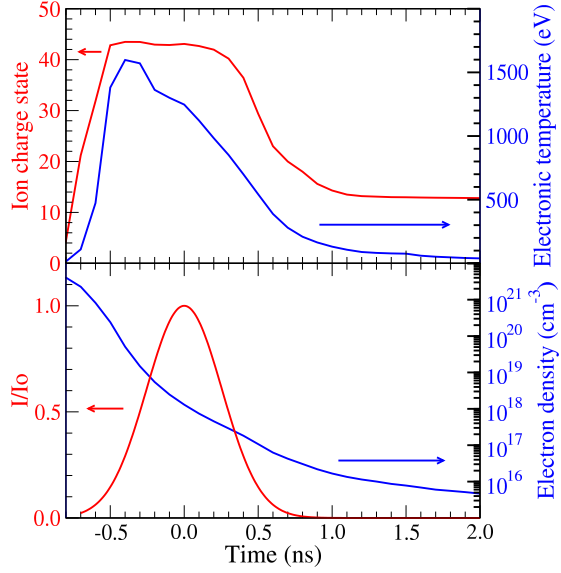}}
\caption{\label{fig:fig9} Electronic temperature, mean ion charge state (top), laser intensity and electron density (bottom) evolutions with time, calculated with Troll in the grid element at the frontier between plasma and vacuum.}
\end{figure}

The time needed for the double layer formation is calculated using the XooPIC  Particle-In-Cell code \cite{verboncoeur1995object} as a function of electron density considering a plasma of Ta$^{20+}$ at the maximum electron temperature reached in the plasma at its frontier with vacuum: 1500 eV. At $t=0$, the plasma is neutral and is localized in one region of the simulation box, the remaining part being vacuum. 
The cell size is adjusted depending on the initial plasma density to ensure numerical stability governed by the Debye length. The time step is fixed at one tenth of the plasma wave period. Electrons expand in vacuum and charge separation occurs in a region where a longitudinal electric field accelerates the ions. The ion energy distribution evolves with calculation time and reaches the steady shape described by Mora$^\prime$s function with the expected $E_0$ value given in Eq.(5), after a given time. The time needed for the transient regime to be gone will be considered in the following as the time needed for the double layer formation. For an electron density of $10^{15}$ e.cm$^{-3}$ and a temperature of 1500 eV, the double layer is created in about 400 ps, $\it i.e.$ a duration of the order of the typical time of plasma evolution ($\sim$ 500 ps). This duration typically depends on the electron density as $n_e^{-1/2}$ as reported in Fig.12(a). Therefore the formation of a double layer efficient for the acceleration of the highest charge state ions requires electron densities larger than $10^{15}$ e.cm$^{-3}$. 

Figure 12(b) shows the evolution of the Debye length as a function of the electron density at the electron temperature of 1500 eV. The maximum value of the plasma gradient measured between $10^{18}$ e.cm$^{-3}$ and $10^{20}$ e.cm$^{-3}$ (see Fig.9) is also reported. We can see that it is much higher than the Debye length. The formation of a double layer thus requires electron densities smaller than $10^{18}$ e.cm$^{-3}$. We can imagine that at the frontier between plasma and vacuum, the density falls to zero in a very thin zone leading to a strong reduction of the plasma gradient at the interface reaching values compatible with the typical values calculated for the Debye length, $\it i.e.$ a few $\mu$m. The conditions for the formation of a double layer could then be achieved in such a specific zone around the plasma-vacuum interface characterized by an electron density ranging from $10^{15}$ to  $10^{18}$ e.cm$^{-3}$. 

\begin{figure}[h]
{\includegraphics[width=7.5cm,trim=0 10 250 0,clip=true]{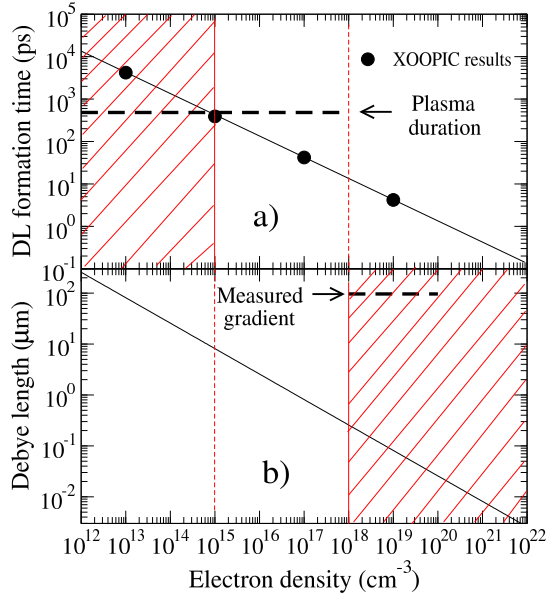}}
\caption{\label{fig:fig12new} a) Double layer (DL) formation time, b) Debye length as a function of electron density in a monocharge plasma of Ta$^{20+}$ at an electron temperature of 1500 eV. The double layer formation times have been calculated with XooPIC codes and fitted with a $n_e^{-1/2}$ function. The plasma gradient measured between $10^{18}$ e.cm$^{-3}$ and $10^{20}$ e.cm$^{-3}$ is reported as well as the plasma duration value estimated from Troll calculations. Theses values define the region of interest for double layer formation (see text).}
\end{figure}

Finally, the PIC simulations show that, at a density of $10^{15}$ e.cm$^{-3}$ and a temperature of 1500 eV, electrons leading to the charge separation are spread over a distance of 1 mm after 200 ps. These electrons are diluted in vacuum and are not experiencing any collisions. Under such conditions, these electrons do not readjust their spatial and velocity distributions 
throughout the plasma cooling stage, leading to a decoupling between the dynamics of the diluted double layer and the evolution with time of the plasma temperature. Therefore the acceleration potential does not strongly evolve with time compared to the evolution of ion charge state in plasma. Different high charge ions can then be successively accelerated during the plasma cooling phase in the double layer created when the electron temperature was hot.

\subsubsection{Energy distributions in a multispecies plasma}

To calculate the energy distributions of the high charge state Ta ions, instead of considering a plasma composition evolving with time, we have treated a simpler system in which a multispecies tantalum plasma with stationnary composition of ions expands in vacuum at a given electron temperature. Ta$^{15+}$, Ta$^{20+}$, Ta$^{25+}$, Ta$^{30+}$ and Ta$^{35+}$ ions are considered in this system with the asymptotic ionic fraction given in Fig. 5. With this composition, the mean charge is about 18+. Calculations are performed as a function of the electron temperature, over a range between 400 eV and 1600 eV. The plasma density has been arbitrarily adjusted to a few $10^{15}$ e.cm$^{-3}$ to set the Debye length at $\lambda_D$=5 $\mu$m for each simulation. Similar studies have been done at higher electron densities without changing the conclusions that will be presented in the following. Typical structures of the electron and ion distributions are shown in Fig. 13(a) as a function of x/$\lambda_D$ at the electron temperature of 1200 eV. Results are reported at $t=250$ ps after the energy distributions have reached their steady shape. Electrons have already moved more than 750 $\mu$m away from their initial position (x$\le$0) while the position of the Ta ion fronts depends on the ion charge state. They can be found up to twenty Debye lengths ($\sim$100 $\mu$m) from their initial position. Figure 13(b) shows the resulting charge separation with two distinct regions: a positive layer on the left of the ion fronts and a negative layer due to the electron cloud on the right. This double layer produces the electric field in which ions experience acceleration at energies depending on their charge states. 

\begin{figure}[h]
{\includegraphics[width=8.cm,trim=0 0 220 0,clip=true]{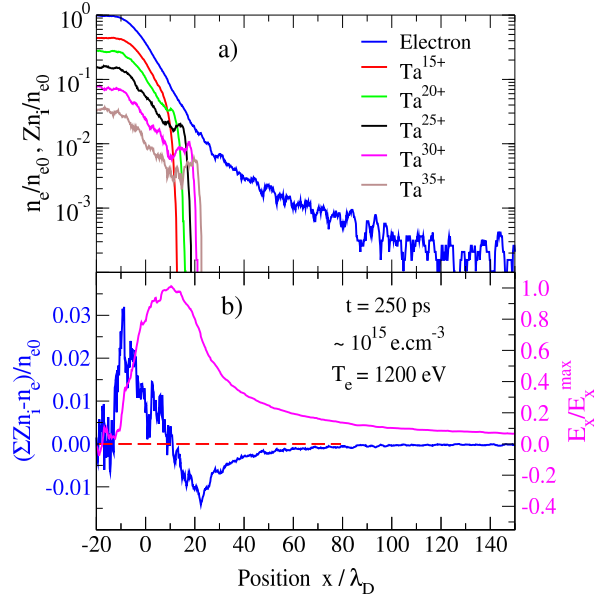}}
\caption{\label{fig:fig10} a) Structure of the ion front at t=250 ps, for a plasma density at about  $10^{15}$ e.cm$^{-3}$ and an electron temperature at 1200 eV. b) Charge separation and longitudinal electric field profile in the same plasma conditions. At t=0, the interface between plasma and vacuum is set at x/$\lambda_D$=0.}
\end{figure}

The obtained energy distributions are reported on Fig.14 for three different charge states. 
They are calculated in the laboratory frame taking the additional velocity $v_B$ of the plasma frontier (corresponding to $E_B$=400 keV) into account as described in section III. Highly charged ions reach higher energies than lowly charged ions and the modified Mora$^\prime$s function defined in Eq.(6) well fits their spectra.  The $E_0$ value strongly depends on the ion charge state and increases from $E_0$= 7$\pm$5 keV for Ta$^{15+}$ to $E_0$=83$\pm$6 keV for Ta$^{35+}$ in these examples. 

\begin{figure}[h]
{\includegraphics[width=6.cm,trim=0 0 270 75,clip=true]{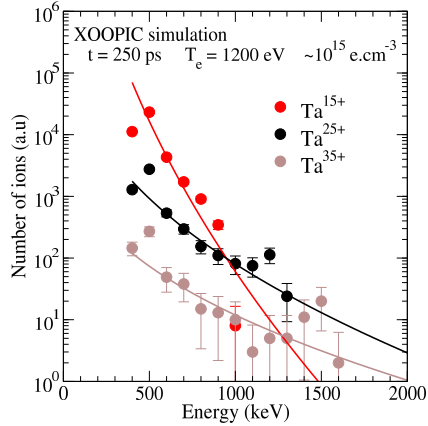}}
\caption{\label{fig:fig11} Calculated energy distributions of Ta$^{15+}$, Ta$^{25+}$ and Ta$^{35+}$ ions in the same plasma conditions as in Fig. 13. Data are fitted with the modified Mora$^\prime$s function defined in Eq.(6).}
\end{figure}

Figure 15 shows the evolution of $E_0$ with the charge state in order to compare experimental and calculated values for different electron temperatures. For each simulation, the energy distributions have been extracted after waiting the duration required for the double layer formation. Agreement between experiment and simulation is observed for electron temperatures around 1200 -1600 eV corresponding to the highest temperatures reached in the plasma periphery. This strengthens the assumption of the formation of a double layer when the plasma is hot leading to the creation of an electric field in which ions are then accelerated throughout the plasma cooling. However, the energy $E_0$ does not linearly depend on the ion charge state as predicted by Mora for a plasma with ions at a single charge number $Q$. Moreover the application of Mora$^\prime$s relation $E_0$=$QkT_e$ gives $E_0$= 42 keV for $Q$=35+ and $T_e$= 1200 eV. This value is about 1.5 - 2 times lower than both the values extracted from experimental data and calculated with XooPIC. In a single charge plasma, Mora showed that the electric field at the ion front scales as $\sqrt{\frac{T_e}{Q}}$ \cite{mora2003plasma}. In the multispecies plasma considered in this XooPIC simulation, the average charge is about 18+ and the electric field could be about $\sqrt{2}$ times larger than the one that would be reached in a plasma only composed of Ta$^{35+}$ ions at the same electron temperature. Such a dependance of the electric field  could explain the high $E_0$ values measured in this work for the highest charge state.

\begin{figure}[h]
{\includegraphics[width=8.cm,trim=0 0 190 70,clip=true]{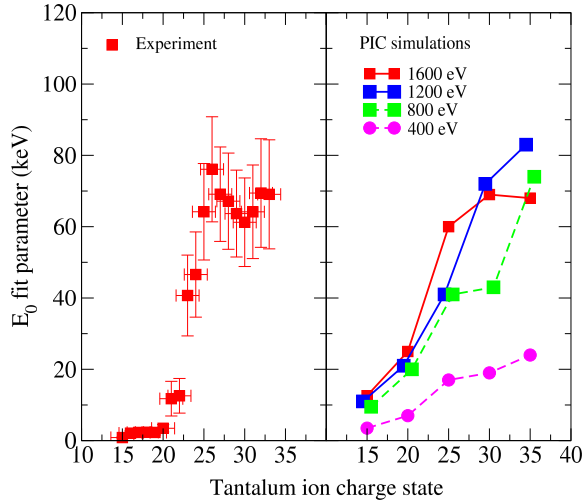}}
\caption{\label{fig:fig12} Evolution of E$_0$ fit parameter as a function of the tantalum ion charge state. Left: experimental value. Right: results of XooPIC simulations for double layers at various electron temperatures.}
\end{figure}

\section{Conclusion}

Energy distributions of ions produced in a nanosecond laser-heated tantalum plasma have been measured far from the target for different charge states. After fitting the energy distributions with MBC functions, these ions can be classified into two components, a ``thermal'' or an ``electric'' one. Under fluid approximation, Troll calculations reproduce the energy distributions of ions in the ``thermal'' component. The ``electric'' component can also be fitted with Mora$^\prime$s function considering a parameter $E_0$  increasing with the ion charge state. This behavior can be reproduced with a kinetic approach in which a double layer is formed at the periphery between plasma and vacuum. This double layer, developed when the electron temperature is at its highest value, produces an electric field in which ions are accelerated throughout the plasma cooling. 

In this scheme and according to the amount of tantalum ions in the ``electric'' component, only nuclei contained in the $\sim$20 first nm of the target are accelerated in this double layer. The remaining part of the nanosecond laser-heated plasma expands with a fluid behavior and experiences important recombination rate leading to the ``thermal'' component observed far from the target. Additional experiments could be interesting to perform in order to confirm and constraint this scheme. One possibility would be to study the ion energy distributions when shooting a thin layer (10 -20 nm thick) of atoms with atomic number $Z_1$ deposited on a thick target with atoms of a different atomic number $Z_2$. This atomic number must be close to $Z_1$ (gold and tantalum for instance) in order not to modify the main bulk hydrodynamics. The temperature and density should therefore be continuous at the interface between both elements and we think that a single double layer should occur at the interface between the target and vacuum. For the adequate thickness of the thin layer, ions with $Z_1$ atomic number would be accelerated in the ``electric'' component whereas ions with $Z_2$ would be present in the ``thermal'' one. 

Such properties are important to take into account for dimensioning future experiments of nuclear physics where high charge state ions are strongly involved and must survive far from the target. In the framework of nuclear excitation in nanosecond laser-heated $^{201}$Hg plasma, this study shows that only the first 20 - 30 nm of the target would be of interest to produce high charge state ions in which nuclei may still be excited far from the target. Considering a focal spot diameter of 20 $\mu$m, Hg nuclei excited in the plasma may be searched for in about 5\textperthousand$ $ of the whole plasma volume, $\it i.e.$ in only 10$^{12}$ ions with charge states higher than 30+. The number of excited nuclei far from the target may therefore be strongly reduced compared to the few millions nuclei expected to be excited in the whole plasma during the laser irradiation. More detailed calculations coupling codes describing the hydrodynamical expansion of plasma and nuclear excitation properties in plasma are currently in progress to extract a quantitative dimensioning of the expected excited nuclei detectable far from the target considering this constraint.

\section*{Acknowledgements}
We thank the staff of the ELFIE facility at Laboratoire pour l$^{\prime}$Utilisation des Lasers Intenses for their technical assistance in running the experiment.
Fruitful discussions with Laurent Gremillet and Paul Edouart Masson-Laborde are kindly acknowledged.
This work was supported by the contract No. 2015-1R60402 of the R\'{e}gion Aquitaine and the IN2P3/CNRS.

\bibliographystyle{apsrev4-1}
\bibliography{version_2018_11_16}

\end{document}